# Early Stopping for Interleaver Recovering of Turbo Codes

Peidong Yu, Hua Peng, Jing Li

Zhengzhou Institute of Information Science and Technology

Zhengzhou, China

yupeidong1234@foxmail.com

## I. Introduction

Parameter recovering for channel codes means to find out coding parameters which are used by the transmitter and are not known to the receiver, through analyzing the received data. The final purpose is to decode the data and retrieve the information. It is important in applications such as *adaptive modulation and coding* and *cognitive radio*. Research has been carried out for different classes of codes. For example, the problem of BCH codes is studied in [1]; different algorithms for convolutional codes are developed in [2], [3]; and methods for code recognition within a candidate set are discussed in [4]~[6] for codes including LDPC ones. For turbo codes, parameters should be recovered for both the convolutional encoders and the interleaver. In this paper, we focus on the problem of recovering the interleaver of turbo codes.

General recovering algorithms [7]-[10] assume random block interleavers, and no prior knowledge about the interleaver structure is needed. Such algorithms usually recover the parameters incrementally from the first one to the last one. Parity-check equations of the coded bits are checked in [7] for deciding each of the parameters. In [8]~[10], the problem is solved through decoding the received data. A hard decoding algorithm is used in [8], and BCJR decoding [11] which uses soft data is exploited in [9], [10]. For each parameter, its probability conditioned on previously recovered parameters and the received data is calculated and maximized in [10], using the forward pass of BCJR decoding. This algorithm given by [10] (we call it *Tillich algorithm* in this paper) is deemed as optimal and has the best performance among existing ones.

However, for Tillich algorithm, the result of each parameter relies on all previous results. This leads to the problem that once some errors occur to the algorithm in the halfway, the results of subsequent parameters may all be wrong. In such a case, there could be lots of wasted computation, since the computational complexity for recovering each parameter is rather considerable and the length of an interleaver is often not small. What's worse, a large number of incorrect interleaver parameters will seriously deteriorate the decoding performance of the turbo code. Early stopping is a technique used in iterative decoding to help avoid useless iterations [12], [13]. We propose that such technique can be used in interleaver recovering, so that not only useless computation, but also incorrect results that will harm the decoding, can be largely avoided. This paper develops an effective and efficient early stopping method for Tillich algorithm, based on the detailed analysis of the algorithm.

The rest of this paper is organized as follows. Section II formulates the recovering problem and gives a description of Tillich algorithm. In section III, firstly the failure property of the algorithm is analyzed, and then an early stopping method is designed making use of the logarithmic probabilities calculated in each iteration of Tillich algorithm, and finally the thresholds are selected through theoretical derivation. Monte Carlo simulation results are shown in section IV. And section V gives the conclusion.

## II. Problem Formulation and Solution

Fig. 1 shows a typical rate 1/3 parallel turbo code encoder with its three outputs passed through a channel. The interleaver is denoted by $\pi = [\pi(1), \cdots, \pi(K)]$, which is of length $K$. And we have $\{\pi(i) \mid i = 1, \cdots, K\} = \{1, \cdots, K\}$. Information sequence $\boldsymbol{u} = (u_1, \cdots, u_K)$ is permuted by the interleaver to be $\boldsymbol{u}^\pi = (u_1^\pi, \cdots, u_K^\pi)$, which satisfies $u_i^\pi = u_{\pi(i)}$ for $i = 1, \cdots, K$, where $u_i \in \text{GF}(2)$ is the information bit. Information sequence $\boldsymbol{u}$ and coded sequences $\boldsymbol{v}$ and $\boldsymbol{w}$ (both of length $K$) are assumed to be modulated using binary phase-shift keying (BPSK) and transmitted through an additive white Gaussian noise (AWGN) channel. Their corresponding received sequences are denoted by $\boldsymbol{x}$, $\boldsymbol{y}$ and $\boldsymbol{z}$ respectively. The interleaver recovering problem of turbo codes means to recover all parameters $\pi(i)$ using these received data. For simplicity, existing algorithms assume that convolutional encoder 2 of the turbo code is already recovered and it always starts from the all-zero state, and the received data are already partitioned into three sequences correctly.

Suppose that M blocks of data are received and denoted by $\{x_s, y_s, z_s : s = 1, \cdots, M\}$, where $x_s = (x_{s,1}, \cdots, x_{s,K})$, $y_s$ and $z_s$ are represented in the same way. Tillich algorithm [10] gives an optimal estimation of the interleaver using the data set $\{x_s, z_s : s = 1, \cdots, M\}$. It recovers $\pi(1), \cdots, \pi(K)$ iteratively and incrementally, with one parameter recovered in each *iteration*.

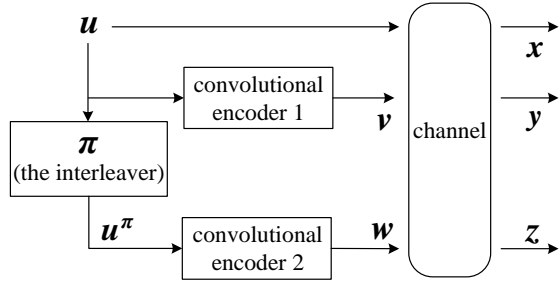

Fig. 1. A typical turbo code encoder and its three outputs.

Consider the *i*th iteration of the algorithm, where the first $(i-1)$ parameters have already been recovered and the results are denoted by $\hat{\pi}(1), \cdots, \hat{\pi}(i-1)$. Using the *s*th block of data, the conditional probability $\Pr(e_{s,i-1} = \alpha \mid x_{s,\hat{\pi}(1)}, \cdots, x_{s,\hat{\pi}(i-1)}, z_{s,1}, \cdots, z_{s,i-1})$, denoted by $h_{s,i-1}(\alpha)$, that the current state $e_{s,i-1}$ of convolutional encoder 2 (see Fig. 1) equals $\alpha$ can be calculated by the forward pass of BCJR decoding. Then the probability $q_{s,i,j}$ that $\pi(i)$ equals $j$ ( $j = 1, \cdots, K$ ) for the *s*th block of data is given by

$$q_{s,i,j} = \frac{\gamma}{\Pr(x_{s,j})} \sum_{a=0,1} \left( \Pr(x_{s,j} \mid u_{s,j} = a) \cdot \sum_{\alpha} \Pr(z_{s,i} \mid w_{s,i} = b) \cdot h_{s,i-1}(\alpha) \right) \quad (1)$$

where $\gamma$ is a normalizing constant, $b$ is the output bit of convolutional encoder 2 when the input bit is $a$ and the state is $\alpha$. Finally, using all $M$ blocks of data, the total probability $p_{i,j}$ that $\pi(i)$ equals $j$ ( $j = 1, \cdots, K$ ) is

$$p_{i,j} = \lambda \prod_{s=1}^{M} q_{s,i,j} \quad (2)$$

where $\lambda$ is a normalizing constant. Thus, the recovered result $\hat{\pi}(i)$ for parameter $\pi(i)$ is

$$\hat{\pi}(i) = \arg\max_{j=1,\cdots,K} (p_{i,j}) \quad (3)$$

The algorithm continues till $K$ iterations are carried out and the whole interleaver is recovered.

We consider that log-domain BCJR decoding [11] is adopted. For each iteration of Tillich algorithm, times of addition and multiplication are both of order $O(MK)$, and times of exponential operations are of order $O(NM)$, where $N$ is the total number of states of convolutional encoder 2 of the turbo code. So, the total times of addition and multiplication needed for the algorithm are $O(MK^2)$, and that of exponential operations are $O(NMK)$.

## III. Early Stopping for Tillich Algorithm
### A. Failure Analysis of the Algorithm

According to (1), the result $\hat{\pi}(i)$ of Tillich algorithm for the *i*th parameter $\pi(i)$ relies on previous results $\hat{\pi}(1), \cdots, \hat{\pi}(i-1)$. More precisely, it relies on the estimated state distributions $h_{s,i-1}(\alpha)$, $s = 1, \cdots, M$, of convolutional encoder 2. If the accuracy of these estimated $h_{s,i-1}(\alpha)$ is low enough, the result $\hat{\pi}(i)$ will be wrong, which leads to even worse accuracy for $h_{s,i}(\alpha)$. Consequently, the algorithm keeps producing wrong results till the end. We say that the algorithm has *failed* if it has fallen into such an *error state*. And when it works correctly, we say that it is in the *correct state*.

Once Tillich algorithm fails, the probability for it to return to the correct state, denoted by $P_{\text{return}}$, is very small. For this, we only give a rough analysis. Assume that after the algorithm has failed, it gives totally random results (an explanation is found in section III-B). Thus the correct recovering probability for each parameter is $1/K$. According to the property of BCJR decoding, in order to make the convolutional encoder state estimation return to the correct state for all blocks of received data, generally $d$ continuous interleaver parameters need to be correctly recovered, where $d$ is the memory depth of convolutional encoder 2. So we have $P_{\text{return}} < K/K^d = 1/K^{d-1}$. For example, when $K=100$ and $d=3$, we have $P_{\text{return}} < 10^{-4}$. This is also checked by simulations, where the algorithm is regarded to have failed if $d$ continuous wrong results are observed, and it is regarded to have returned to the correct state if $d$ continuous correct results appear after it has failed. Different sets of values of $K$, $d$, $M$ and signal-to-noise ratio (SNR) are considered. For each set of the values, 1000 cases where the algorithm fails are examined. And we have not observed any case where the algorithm has returned to the correct state.

Therefore, it is necessary to design an early stopping method for Tillich algorithm to make it stop in time after it fails. By early stopping, useless iterations and wrong results can be largely avoided. And also, the successfulness of the algorithm can be monitored: it is not successful if it is stopped by the proposed method; otherwise, it can be viewed as successful.

*B. Proposed Early Stopping Method*

To ensure numerical stability of Tillich algorithm, the log-domain BCJR decoding should be adopted. And for the $i$th iteration of the algorithm, probability given by (2) is replaced by its logarithmic form, which is

$$l_{i,j} \triangleq \log(p_{i,j}) = \log(\lambda) + \sum_{s=1}^{M} \log(q_{s,i,j}) \tag{4}$$

for $j = 1, \cdots, K$. The result $\hat{\pi}(i)$ for the $i$th parameter equals the $j$ which maximizes $l_{i,j}$.

For a given $i$, when computing $l_{i,j}$ using (4) for different values of $j$, the received data $x_{s,\hat{\pi}(k)}$, $z_{s,k}$ and $z_{s,i}$ ($s = 1, \cdots, M$; $k = 1, \cdots, i-1$) are all fixed values and can be viewed as constants. Each $l_{i,j}$ only depends on variables $x_{1,j}, \cdots, x_{M,j}$, denoted by $\mathbf{x}_{(j)}$. Since the independence among $\mathbf{x}_{(1)}, \cdots, \mathbf{x}_{(K)}$, variables $l_{i,1}, \cdots, l_{i,K}$ can be viewed as independent of each other. For a given $i$ and a fixed $j$, $\log(q_{1,i,j}), \cdots, \log(q_{M,i,j})$ are independent and identically distributed (i.i.d.), since $\log(q_{s,i,j})$ only depends on the $s$th block of received data. According the central limit theorem, $l_{i,j}$ given by (4) tends to follow Gaussian distribution when $M$ is large.

By the above analysis, we make the following two reasonable assumptions. Firstly, variables $l_{i,1}, \cdots, l_{i,K}$ are independent of each other; and secondly, each of them is Gaussian distributed.

At the $i$th iteration, Tillich algorithm either has failed or has not. If it has already failed, there exists $r > 1$ such that $\forall j$, there is no detectable coding constraint that holds for all $M$ sequence pairs, i.e. $(x_{s,\hat{\pi}(i-r)}, \cdots, x_{s,\hat{\pi}(i-1)}, x_{s,j})$ and $(z_{s,i-r}, \cdots, z_{s,i-1}, z_{s,i})$ for $s = 1, \cdots, M$ (so that values $l_{i,j}$ are too inaccurate to produce the correct result). In this case, variables $l_{i,1}, \cdots, l_{i,K}$ are i.i.d. (which explains the random results assumption in section III-A) and we denote $l_{i,j} \sim \mathbb{N}(u_0, \sigma_0^2)$, $\forall j$. If the algorithm has not failed, previously recovered parameters must be almost all correct, and thus when $j = \pi(i)$, all sequence pairs $(x_{s,\hat{\pi}(1)}, \cdots, x_{s,\hat{\pi}(i-1)}, x_{s,j})$ and $(z_{s,1}, \cdots, z_{s,i-1}, z_{s,i})$ correspond to true coded sequences; and when $j \neq \pi(i)$, this still holds except that $x_{s,j}$ in each pair is from a random bit. In this case, variables $l_{i,1}, \cdots, l_{i,\pi(i)-1}, l_{i,\pi(i)+1} \cdots, l_{i,K}$ are i.i.d. and we denote $l_{i,j} \sim \mathbb{N}(u_2, \sigma_2^2)$ for $j \neq \pi(i)$, and denote $l_{i,\pi(i)} \sim \mathbb{N}(u_1, \sigma_1^2)$. Obviously, we have $u_1 > u_2 > u_0$ in general.

Thus, to decide whether the algorithm has failed or not, a straightforward method is to use the mean value of $l_{i,1}, \cdots, l_{i,K}$. It would be smaller than a certain threshold if the algorithm has failed. However, such a threshold is difficult to set, since parameters of the Gaussian distributions above are determined by $M$, SNR and parameters of convolutional encoder 2, and are not yet derived. In the following, we propose for Tillich algorithm an effective and efficient early stopping method, for which the thresholds can be set conveniently.

Denote the maximum value in $l_{i,1}, \cdots, l_{i,K}$ by $l_{i,j'}$ which is the $j'$th value. Calculate the mean $u$ and standard deviation $\sigma$ of the rest ($K$-1) values as

$$u = \frac{1}{K-1} \sum_{j=1, j \neq j'}^{K} l_{i,j} \tag{5}$$

$$\sigma = \sqrt{-u^2 + \frac{1}{K-1} \sum_{j=1, j \neq j'}^{K} l_{i,j}^2} \tag{6}$$

Compute the normalized difference $\varepsilon_i$ between $l_{i,j'}$ and $u$, that is

$$\varepsilon_i = (l_{i,j'} - u)/\sigma \tag{7}$$

If the algorithm is in the correct state, there is a large probability that the maximum value $l_{i,j'}$ equals $l_{i,\pi(i)}$

and is outstanding among other values, leading $\varepsilon_i$ to be large; and if the algorithm has failed, all values $l_{i,j}, \forall j$ follow the same Gaussian distribution, resulting in relatively smaller $\varepsilon_i$.

Then the proposed early stopping method is described as: set thresholds $A$ and $B$; if $\varepsilon_i < A$, where $i \in \{1, \cdots, K\}$, holds for $B$ continuous iterations of Tillich algorithm, i.e. $\varepsilon_{i-B+1}, \cdots, \varepsilon_i < A$ happens for any $i$, Tillich algorithm stops; otherwise, the algorithm goes on. Threshold $A$ is a positive real number, and $B$ is an integer. Values for them, as well as the necessity of employing threshold $B$, are discussed in the next subsection.

For each iteration of Tillich algorithm, the proposed method has to compute (5)-(7) and count the comparison result between $\varepsilon_i$ and $A$. This needs only about $K$ times of multiplication and $2K$ times of addition, and the size of extra memory space needed is only of order $O(1)$.

*C. Setting the Thresholds*

Suppose that in a certain implementation of Tillich algorithm, it works in the correct state for the first $i_1$ iterations, and it falls into the error state since the $(i_1+1)$th iteration. When applying the proposed early stopping method and implemented for the same data, it stops at the end of the $i_2$th iteration. Denote $W = i_2 - i_1$ if we have $i_2 > i_1$, which means the algorithm stops after it has failed. And denote $L = i_1 - i_2$ if we have $i_2 \leq i_1$, which implies that the algorithm stops mistakenly before it fails. Thresholds $A$ and $B$ should be such that expectations $E(W)$ and $E(L)$ are both as small as possible.

In the following derivation, we take the results given by (5) and (6) as the true expectation and standard deviation of corresponding variables. This is reasonable when $K$ is large.

To derive $E(W)$, for simplicity, we consider the case where Tillich algorithm is in the error state since the first iteration. The probability $p_0$ that $\varepsilon_i < A$ holds can be easily derived as

$$p_0 = (1 - Q(A))^K \tag{8}$$

where $Q(\alpha) = \int_\alpha^{+\infty} f(x)dx$ with $f(x)$ being probability density function of the standard Gaussian distribution $\mathbb{N}(0,1)$. Denote by $P_0(k)$ the probability that the algorithm stops at the end of the $k$th ($k = 1, 2, 3, \cdots$) iteration. It can be verified that

$$P_0(k) = \begin{cases} 0, & k \leq B-1; \\ p_0^B, & k = B; \\ p_0^B(1-p_0)\left(1 - \sum_{i=1}^{k-B-1} P_0(i)\right), & B+1 \leq k \leq K \end{cases} \tag{9}$$

Thus the expectation $E(W)$ for the considered case is derived as

$$E(W) = \sum_{k=1}^{K} k \cdot P_0(k) \tag{10}$$

This is actually an upper bound (or an asymptotic value) of the true expectation. It may be approached when $P_0(k)$ vanishes fast as $k$ increases or when the algorithm works under low SNR conditions so that it fails quickly.

Next, we derive the expectation $E(L)$. For simplicity, assume $\sigma_1 = \sigma_2 \stackrel{\text{def}}{=} \sigma$ for the above mentioned Gaussian distributions and denote $T = (u_1 - u_2)/\sigma$. When Tillich algorithm works in the correct state, the probability $p_1$ that the current parameter $\pi(i)$ is correctly recovered is

$$p_1 = \Pr\left(l_{i,\pi(i)} > \max_{j \neq \pi(i)} l_{i,j}\right) = \int_{-\infty}^{+\infty}(1 - Q(x+T))^{K-1} f(x) dx \tag{11}$$

where functions $Q(x)$ and $f(x)$ are defined as above. Assume that one single incorrectly recovered parameter will make the algorithm to fail. Denote by $P_1(k)$ the probability that the algorithm fails after the first $k$ ($k = 1, \cdots, K$) parameters have been correctly recovered, and we have

$$P_1(k) = \begin{cases} p_1^k(1-p_1), & k = 1, \cdots, K-1; \\ p_1^K, & k = K \end{cases} \tag{12}$$

Given that $\pi(i)$ is correctly recovered, i.e. the maximum value $l_{i,j'}$ involved in (7) equals $l_{i,\pi(i)}$, the probability $p_2$ that $\varepsilon_i < A$ holds is

$$p_2 = 1 - Q\big((A\sigma + u - u_1)/\sigma\big) = Q(T - A) \tag{13}$$

Provided that the algorithm could stay in the correct state for $k$ iterations before it fails, denote by $P_2(k,i)$ the probability that it is stopped at the end of the $i$th ($i<k$) iteration by the early stopping method. This probability can be derived in the same way as for $P_0(k)$, and is given by

$$P_2(k,i) = \begin{cases} 0, & i \leq B-1; \\ p_2^B, & i = B; \\ p_2^B (1-p_2)\left(1 - \sum_{i'=1}^{i-B-1} P_2(k,i')\right), & B+1 \leq i \leq k \end{cases} \tag{14}$$

Thus the expectation $E(L)$ is

$$E(L) = \sum_{k=1}^{K}\left( P_1(k) \cdot \sum_{i=1}^{k-1}(k-i)\cdot P_2(k,i) \right) \tag{15}$$

which is a function of $A$, $B$ and $T$. For fixed thresholds $A$ and $B$, we consider the maximum value $E_{\max}(L)$ of $E(L)$, which can be obtained through numerical searching with respect to $T$.

From the above derivation, we see that both thresholds are determined only by the interleaver length $K$, and are not affected by $M$, SNR or the convolutional encoder parameters of the turbo code. Fig. 2 shows how $E(W)$ and $E_{\max}(L)$ change against the thresholds when $K=512$. Values of $E_{\max}(L)$ are multiplied by constant $C=100$ for convenience of displaying. For a fixed $B$, as $A$ grows larger, we have that $E(W)$ turns smaller and $E_{\max}(L)$ grows larger. The thresholds should be such that $E(W)$ and $E_{\max}(L)$ are both small enough at the same time. For small values of $B$ for example $B=1$ or $B=2$, this is difficult to be achieved as is shown in the figure (when $B=1$, the curve of $C \cdot E_{\max}(L)$ is not found because it is almost out of the scope of the figure), which explains that threshold $B$ is necessary for the proposed method. However, $B$ should not be too large since we have $E(W) \geq B$ from (10).

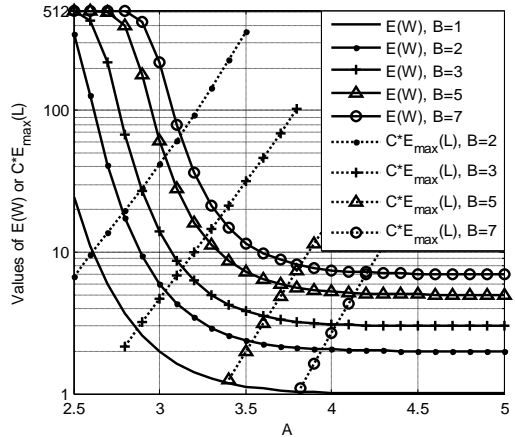

Fig. 2. Values of $E(W)$ and $E_{\max}(L)$ versus thresholds $A$ and $B$ for interleaver length $K=512$.

Given $K$ and the required constraints on $E(W)$ and $E_{\max}(L)$, the thresholds are set to be values that meet the constraints. These values are obtained using (10) and (15) through numerical search, and are stored for future use. Usually, values of the thresholds that meet the constraints are not unique and we can choose at will. And if the constraints are too strict to be met, they should be relaxed till there is a solution for the thresholds. In practice, it is required that $E(W) \ll K$, $E_{\max}(L) \ll 1$. Take $K=512$ for example. Constraints $E(W) < 5$ and $E_{\max}(L) < 0.01$ result in no solution for $A$ and $B$. But if we only need $E(W) < 10$ and $E_{\max}(L) < 0.02$, we can have $5 \leq B \leq 9$ and for $B=5$ for instance, we have $3.34 \leq A \leq 3.50$.

When $B$ is fixed, larger $E(W)$ means smaller $E_{\max}(L)$. So, considering that $E_{\max}(L)$ is less convenient than $E(W)$ to be evaluated, we can also set the thresholds by choosing proper values for $B$ and $E(W)$ (thus $A$ is also fixed). As an example, Table 1 shows values of $A$ and $E_{\max}(L)$ for different values of $K$, given $B=5$ and $E(W) = 7.5$. As is shown, we have $E_{\max}(L) < 0.02$ for any $K \geq 512$; and $E_{\max}(L)$ decreases as $K$ increases. And $E_{\max}(L)$ can still be reduced by increasing $E(W)$, as long as $E(W)$ is kept much smaller than $K$.

Table 1. Values of threshold $A$ and $E_{\max}(L)$ for different interleaver lengths $K$, given threshold $B=5$ and $E(W)=7.5$.

| $K$ | 512 | 1024 | 2048 | 4096 | 8192 | 16384 |
|---|---|---|---|---|---|---|
| $A$ | 3.48 | 3.66 | 3.83 | 4.00 | 4.16 | 4.32 |
| $100E_{\max}(L)$ | 1.82 | 1.65 | 1.50 | 1.43 | 1.34 | 1.29 |

## IV. Simulation Results

In this section, we check performance of the proposed early stopping method by simulations. For Tillich algorithm, [10] considers correct probability of the whole interleaver. In practice, even a partially recovered interleaver will benefit the turbo decoding. Thus, we define the correct probability $P_C$ by $K_C/K$, where $K_C$ is the average number of correctly recovered parameters for each time the algorithm is implemented. If an iteration of Tillich algorithm yields an incorrect result, we call it a *wasted iteration*. The expectation $E(W)$ is measured by the average number of incorrectly recovered parameters, which equals the average number of wasted iterations. Since computational complexity of each iteration is fixed, $E(W)$ measures the wasted computation of the algorithm. Both quantities are obtained through Monte Carlo trials.

Fig. 3 shows $P_C$ and $E(W)$ of Tillich algorithm as functions of SNR, for interleavers of length $K=512$. The recursive encoder with generator polynomial $(1+D+D^2+D^4)/(1+D^3+D^4)$, which is of memory depth $d=4$, is adopted as convolutional encoder 2 (see Fig. 1) of the turbo code. Different numbers $M$ of data blocks are considered. Thresholds are set to be $A=3.48$ and $B=5$ according to Table 1. In the figures, "NES" denotes simulation results when *no early stopping* is applied to the algorithm, and "ES" denotes that when the proposed early stopping method is applied. Fig. 3(a) shows that early stopping leads to no noticeable degradation in correct probability. In Fig. 3(b), we see that, without early stopping, the average number of wasted iterations increases to approach $K$, as SNR drops; while with early stopping, this number keeps much smaller and it stops increasing and stays around the value 7.5 at low SNRs ($7.5 \ll K$). Varying the generator polynomial of convolutional encoder 2, results do not change. And for different $K$ (thresholds are taken from Table 1 accordingly), similar results are observed. These verify the effectiveness of the proposed method and the provided thresholds, which are constants given the value of $K$ and not affected by other factors.

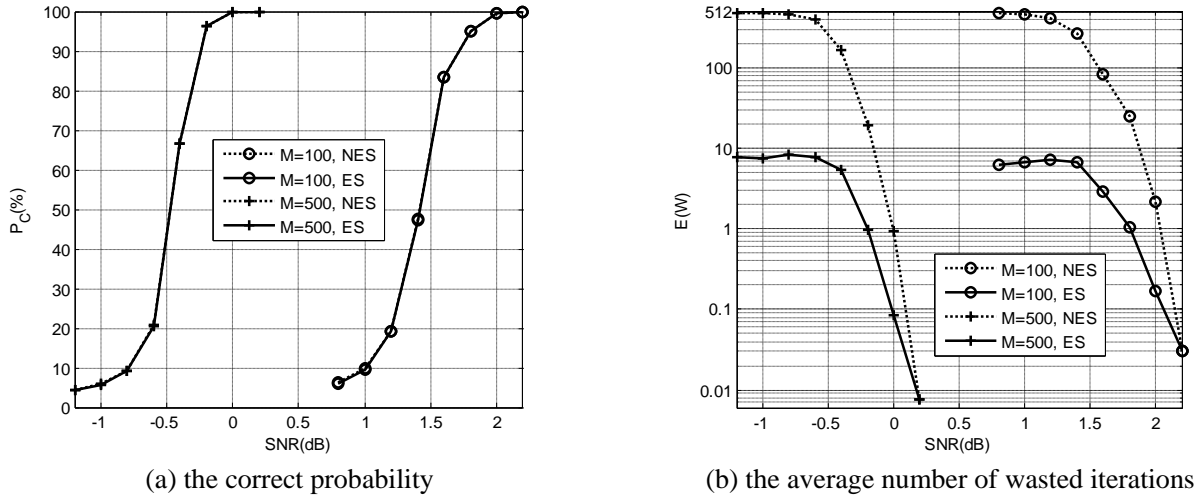

(a) the correct probability  (b) the average number of wasted iterations

Fig. 3. Effect of the proposed early stopping method on performance of Tillich algorithm, for interleavers of length $K=512$. Thresholds for the method are set according to Table 1.

Fig. 4 gives performance of the proposed method for different values of thresholds $A$ and $B$. Interleavers of length $K=1024$ are considered and $M=100$ blocks of data are used. The thresholds are firstly taken from Table 1, and then $B$ is decreased by 1 or increased by 2 while $A$ is not changed, and finally $A$ is decreased by 0.5 or increased by 0.3 while $B$ is not changed. Denote by $P_{C,ES}$ and $P_{C,NES}$ respectively the correct probabilities of Tillich algorithm with or without early stopping. Thus $\Delta P_C = P_{C,NES} - P_{C,ES}$ denotes the reduction in correct probability since applying early stopping. Results for the algorithm without early stopping are also shown in the figures.

For thresholds from Table 1, the maximum value of $\Delta P_C$ is about 0.1% when SNR=1.75dB, as shown in Fig. 4(a). This means in each time of recovering, the maximum reduced number of correctly recovered parameters is about $E_{\max}(L) \approx K \times 0.1\% \approx 1$. This simulated result of $E_{\max}(L)$ is larger than that given in Table 1 mainly for two reasons. Firstly, Tillich algorithm often shows some unpredictable instability at the critical moment when it is to fail. Secondly, there should still be a correct probability of $1/K$ after the algorithm has failed if it is not stopped. In practice, such unpredictable correctly recovered parameters are obviously useless. As shown in Fig. 4(b), $E(W)$ is not larger than 7.5 for thresholds from Table 1, which coincide with the theoretical result.

By varying the thresholds, the maximum value of $\Delta P_C$ increases when $E(W)$ decreases, and vice versa, which also coincide with the theoretical analysis. As shown in Fig. 4, both quantities are not sensitive to the thresholds. For example, when $A$ decreases by 0.5 or $B$ increases by 2 with respect to their original values from Table 1, values of $\Delta P_C$ decrease (which means even negligible degradation in the correct probability), while $E(W)$ still keeps smaller than 10 or 25 respectively ($25 \ll K$).

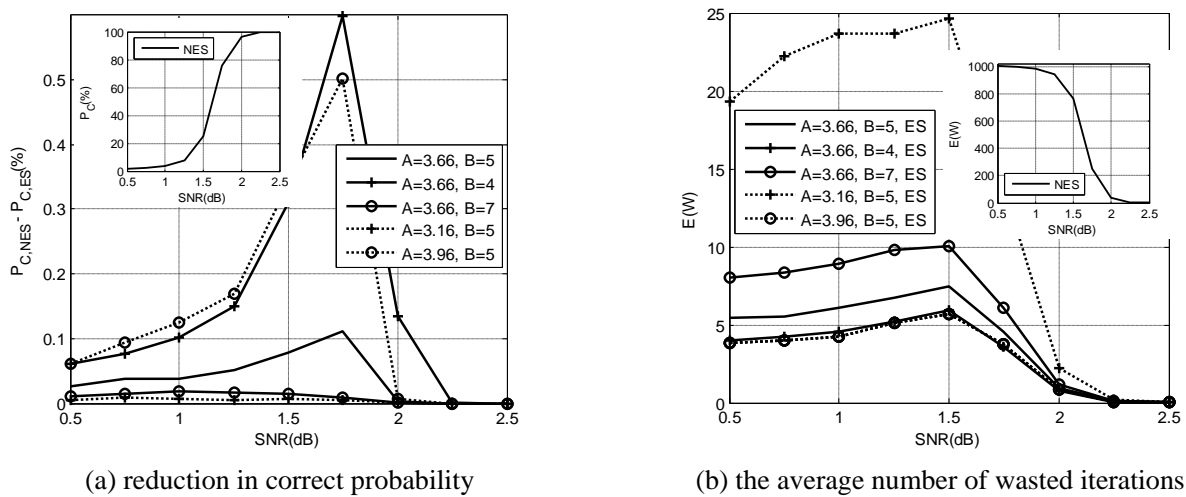

(a) reduction in correct probability    (b) the average number of wasted iterations

Fig. 4.  Impact of values taken for thresholds of the proposed early stopping method on performance of the method, for interleavers of length $K$=1024.

## V. Conclusion

This paper deals with the interleaver recovering problem of turbo codes. An early stopping method is designed for the existing optimal algorithm. The method needs two thresholds, which are both determined only by the interleaver length and are obtained through theoretical derivation. Simulations show that, while keeping negligible degradation in the correct recovering probability, the proposed method is able to stop the algorithm in time after it has failed. Lots of useless iterations and incorrectly recovered parameters are thus avoided, especially when the algorithm works under relatively low SNR conditions.